\documentclass[journal=jacsat,manuscript=article,layout=twocolumn]{achemso}

\usepackage{bm}
\usepackage{color}
\usepackage{soul}
\usepackage{amsmath}
\usepackage{amssymb}
\usepackage{booktabs}

\author{Kenan Song}
\affiliation
{Catalan Institute of Nanoscience and Nanotechnology (ICN2), CSIC and BIST, Campus UAB, 08193 Barcelona, Spain}
\author{David Soriano}
\affiliation
{QuantaLab International Iberian Nanotechnology Laboratory (INL), Av. Mestre Jose Veiga, 4715-330 Braga, Portugal}
\author{Aron W. Cummings}
\author{Roberto Robles}
\author{Pablo Ordej\'on}
\affiliation
{Catalan Institute of Nanoscience and Nanotechnology (ICN2), CSIC and BIST, Campus UAB, 08193 Barcelona, Spain}
\author{Stephan Roche}
\email{stephan.roche@icn2.cat}
\affiliation
{Catalan Institute of Nanoscience and Nanotechnology (ICN2), CSIC and BIST, Campus UAB, 08193 Barcelona, Spain}
\alsoaffiliation
{ICREA - Instituci\'o Catalana de Recerca i Estudis Avan\c{c}ats, 08010 Barcelona, Spain}

\title{Spin Proximity Effects in Graphene/Topological Insulator Heterostructures}

\begin{document}

\begin{abstract}
Enhancing the spin-orbit interaction in graphene, via proximity effects with topological insulators, could create a novel 2D system that combines nontrivial spin textures with high electron mobility. In order to engineer practical spintronics applications with such graphene/topological insulator (Gr/TI) heterostructures, an understanding of the hybrid spin-dependent properties is essential. {However to date, despite the large number of experimental studies on Gr/TI heterostructures reporting a great variety of remarkable (spin) transport phenomena, little is known about the true nature of the spin texture of the interface states as well as their role on the measured properties. Here we use {\it ab initio} simulations and tight-binding models to determine the precise spin texture of electronic states in graphene interfaced with a Bi$_2$Se$_3$ topological insulator.  Our calculations predict the emergence of a giant spin lifetime anisotropy in the graphene layer, which should be a measurable hallmark of spin transport in Gr/TI heterostructures, and suggest novel types of spin devices.}

\end{abstract}

\section{Introduction}

Following the discovery of graphene and the large family of van der Waals heterostructures based on two-dimensional materials \cite{Geim2014, Butler2013, Ganesh2015, Novoselov2016, Lin2016}, the spectrum of practical applications harnessing the uniqueness of such materials has grown continuously \cite{Ferrari2015}. In the area of spintronics \cite{Roche2015, Kawakami2015}, the long room-temperature spin lifetime in graphene opens the possibility of large-scale integration of lateral spintronic devices and architectures \cite{Drogeler2014, Drogeler2016, Ingla2016}. Additionally, many recent reports indicate the benefit of using proximity effects to tune the spin properties inside the graphene layer and to engineer devices such as spin field-effect transistors \cite{Hueso2016, Saroj2017}. This provides exciting opportunities in the search for innovative spin manipulation strategies and the development of non-charge-based information processing technologies.  Proximity effects have been studied by combining graphene with magnetic insulators \cite{Leutenantsmeyer2016, Yang2013, Hallal2017, Singh2017}, or by magnifying the spin-orbit coupling (SOC) in graphene through extrinsic chemical functionalization \cite{Avsar2015, Cresti2016, VanTuan2016}. Another route recently proposed is to interface graphene with transition metal dichalcogenides (TMDCs) \cite{Schmidt2016, Ge2017} such as WS$_2$ or WSe$_2$, which leads to phenomena such as weak antilocalization (WAL) \cite{Wang2015, Wang2016, Yang2016} or large nonlocal Hall signals \cite{Avsar2014}. Additionally, the fact that electronic states in Gr/TMDC systems are spin-polarized primarily along the out-of-plane direction \cite{Gmitra2016} results in large spin lifetime anisotropy between in-plane and out-of-plane spin-polarized electrons, which is however weakly energy dependent and therefore not tunable \cite{Cummings2017, Ghiasi2017, Benitez2017}.

Recently, a lot of attention has been paid to heterostructures of graphene and topological insulators (TIs), with reports of anomalous magnetotransport, giant Edelstein effect, and gate-tunable tunneling resistance \cite{Kim2014,Zhang2016,ZhangACS2017,Zalic2017,Rossi2017}, as well as the possible existence of a quantum spin Hall phase \cite{PaengroACS2015, Cao2016}. On the more applied side, the fabrication of broadband photodetectors based on Gr/TI heterostructures has been realized \cite{QiaoACS2015}, as well as the injection of spin-polarized current from an ultrathin Bi$_2$Te$_2$Se nanoplatelet into graphene \cite{VaklinovaNL2016}. TI materials are distinguished by their strong intrinsic SOC, which leads to the formation of a bulk band gap and 2D surface states that host massless Dirac fermions with spin-momentum locking \cite{Bernevig1757, Hsieh2009, Moore2010, Hasan2010,Bercioux2015,Soumyanarayanan2016}. Proximity to a TI leads to a band gap opening and spin-split bands in graphene, as discussed theoretically for the case of Bi$_2$Se$_3$ \cite{Popov2014, Rajput2016}, or for graphene interfaced with Sb$_2$Te$_3$ \cite{Cao2016}.

However, currently there is substantial variability in the literature concerning the precise spin characteristics of Gr/TI systems. Rajput and coworkers measured and calculated a spin splitting of $\sim$80 meV in graphene on Bi$_2$Se$_3$ \cite{Rajput2016}, while Lee {\it et al.} calculated a band gap of up to 20 meV induced in graphene by Bi$_2$Te$_2$Se when all Dirac cones coincided \cite{doi:10.1021/acsnano.5b03821}. Kou {\it et al.} predicted a SOC of $\sim$2 meV induced in graphene sandwiched between two layers of Sb$_2$Te$_3$ \cite{Kou2015418}. They also pointed out, as did Lin {\it et al.} \cite{Lin2017}, the importance of the Kekul\'e distortion on the magnitude of the band gap in graphene. Jin and Jhi reported a TI thickness dependence of SOC induced in graphene by Sb$_2$Te$_3$, and they also hinted at unusual spin textures induced in the graphene bands \cite{PhysRevB.87.075442}. Meanwhile, De Beule {\it et al.} \cite{Beule2017} concluded that the spin texture imprinted on the graphene states should resemble the standard Rashba texture, as also found in Zhang {\it et al.} \cite{ZhangJunhua} 
Overall, these works indicate that TIs clearly induce strong proximity effects, resulting in gap opening and spin splitting of the bands, but the precise nature of the spin texture induced in the graphene layer, and the way to detect it experimentally are crucially lacking. Additionally, such information is not only essential for clarifying how proximity effects between graphene and TIs generate the measured properties, but could also enlarge the possibilities for tailoring spintronics applications.

{In this article, we report fundamental spin transport properties of Gr/TI heterostructures, by performing {\it ab initio} calculations and fitting to tight-binding (TB) models that fully reproduce both the band structure and the spin texture in the graphene layer. Structures with different twist angles between the graphene and the TI are considered, but in all cases a giant spin lifetime anisotropy in the graphene layer, with in-plane spins relaxing much faster than out-of-plane spins is obtained. In the highly commensurate structure, with a twist angle of $30^\circ$, the anisotropy is maximal near the graphene Dirac point, reaching values of tens to hundreds, and decays to $1/2$ at higher energies. Meanwhile, in the larger unit cell, with a twist angle of $0^\circ$, the anisotropy remains high at all Fermi energies and exhibits a strong electron-hole asymmetry. The difference in these behaviors is driven by the dominating SOC terms in each structure, which depend on the specific interface symmetry.  This contrasts with  prior calculations of the spin texture in Gr/TI systems \cite{Beule2017,ZhangJunhua}, which predicted a purely Rashba-like spin texture with an energy-independent anisotropy of 1/2.

Our theoretical predictions could be experimentally confirmed by performing a gate-dependent measurement of the anisotropy as recently achieved in Gr/TMDC samples \cite{Ghiasi2017, Benitez2017, Raes2016, Raes2017}; heterostructures which however do not exhibit any gate-dependence. Differently, Gr/TI allows for a strong gate-dependent anisotropy enabling the fabrication of tunable spin filtering devices, while the in-plane spin-momentum locking could also make it possible to convert charge current to spin current and to control the spin orientation of the current \cite{VaklinovaNL2016}.

\section{Model and Methods}

{\it Ab initio} calculations of the electronic structure of Gr/TI heterostructures were carried out using density functional theory (DFT) \cite{Kohn1965}, implemented in the Vienna Ab initio Simulation Package (VASP) \cite{Kresse1}, with the wave functions expanded in a plane wave basis with an energy cutoff of 600 eV, using the projector augmented wave method \cite{Kresse2}. The PBE form of the generalized gradient approximation \cite{Perdew1997} was used to compute the exchange-correlation energy, and a $24\times 24\times 1$ ($9\times 9\times 1$) k-point mesh for the small (large) unit cell was used together with a convergence criterion of $10^{-6}$ eV. The spin-orbit coupling was included through noncollinear calculations, while the Van der Waals force was accounted for based on the Tkatchenko-Scheffler method \cite{Tkatchenko2009}, and all structures were fully relaxed until forces were smaller than $10^{-2}$ eV/\AA. 

Figure \ref{fig_structure} shows the simulated Gr/TI heterostructures, which contain a Bi$_2$Se$_3$ film and one monolayer of graphene in either a $\sqrt{3} \times \sqrt{3}$ or a $5 \times 5$ supercell, shown in Figs.\ \ref{fig_structure}(a) and (b), respectively. Since the graphene layer is attached to the TI substrate, the minimum-energy lattice constant of bulk Bi$_2$Se$_3$, 4.196 \AA, was adopted. The relaxed crystal structures exhibit a lattice mismatch of less than $3\%$, while the relaxed interlayer spacing between the graphene and TI is larger than 3.5 \AA. For the small unit cell we considered TI films of two different thicknesses, one and six quintuple layers (QLs), as indicated in Fig.\ \ref{fig_structure}(c), while for the large unit cell we only considered a thickness of 1QL. Since our results are qualitatively independent of the number of QLs, we restrict our discussion to the 1QL case, and show the 6QL results in the Supplemental Material. Figure \ref{fig_structure}(a) depicts the hollow configuration, with the top (green) Se atom in the center of a carbon ring. This is the most stable configuration, but when studying the spin texture in this unit cell we have also considered other alignments. Figure \ref{fig_structure}(b) shows the structure of the larger unit cell, which contains a mix of hollow, top, and bridge alignments between C and Se.

\begin{figure}[t]
\includegraphics[width=\columnwidth]{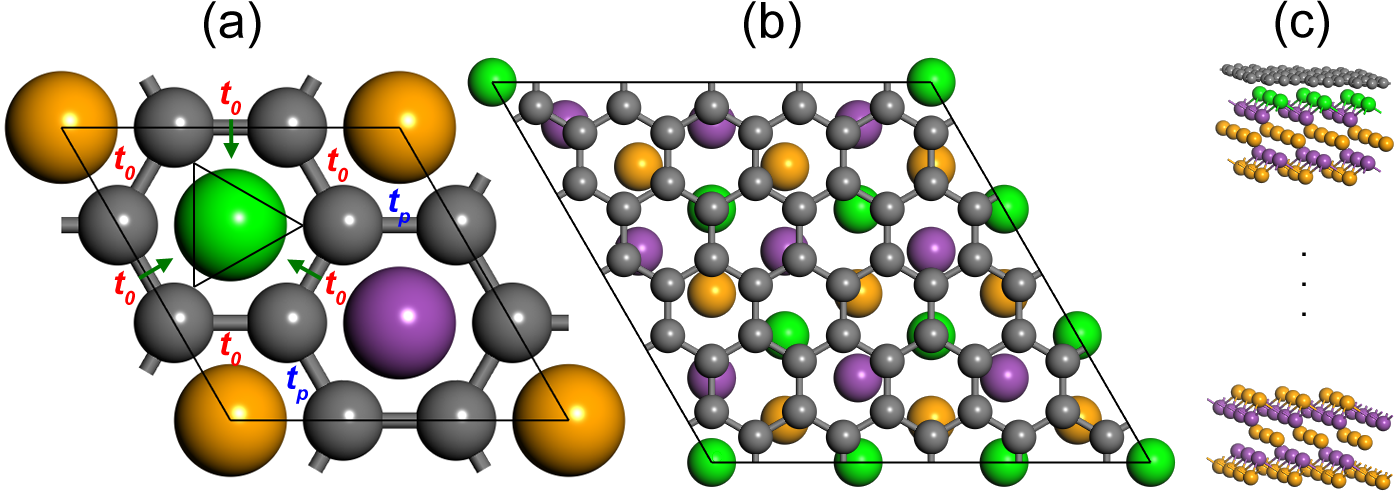}
\caption{Unit cells and TB parameters for the Gr/Bi$_2$Se$_3$ heterostructure. Green atoms are Se in the top TI layer, while purple and orange atoms are Bi and Se on deeper layers. Grey atoms are carbon. (a) In the small unit cell, the hopping within and between carbon rings is denoted by $t_0$ and $t_p$, respectively. The triangle indicates intrinsic SOC within the primary carbon ring. Green arrows indicate the electric fields responsible for in-plane Rashba SOC, and a uniform out-of-plane Rashba SOC is also assumed. (b) The larger unit cell includes a variety of C-Se alignments. For this system we use a TB model originally developed for Gr/TMDC heterostructures \cite{Gmitra2016}. (c) Side view of the structures, indicating a variable number of QLs.}
\label{fig_structure}
\end{figure}

\begin{figure}[t]
\includegraphics[width=\columnwidth]{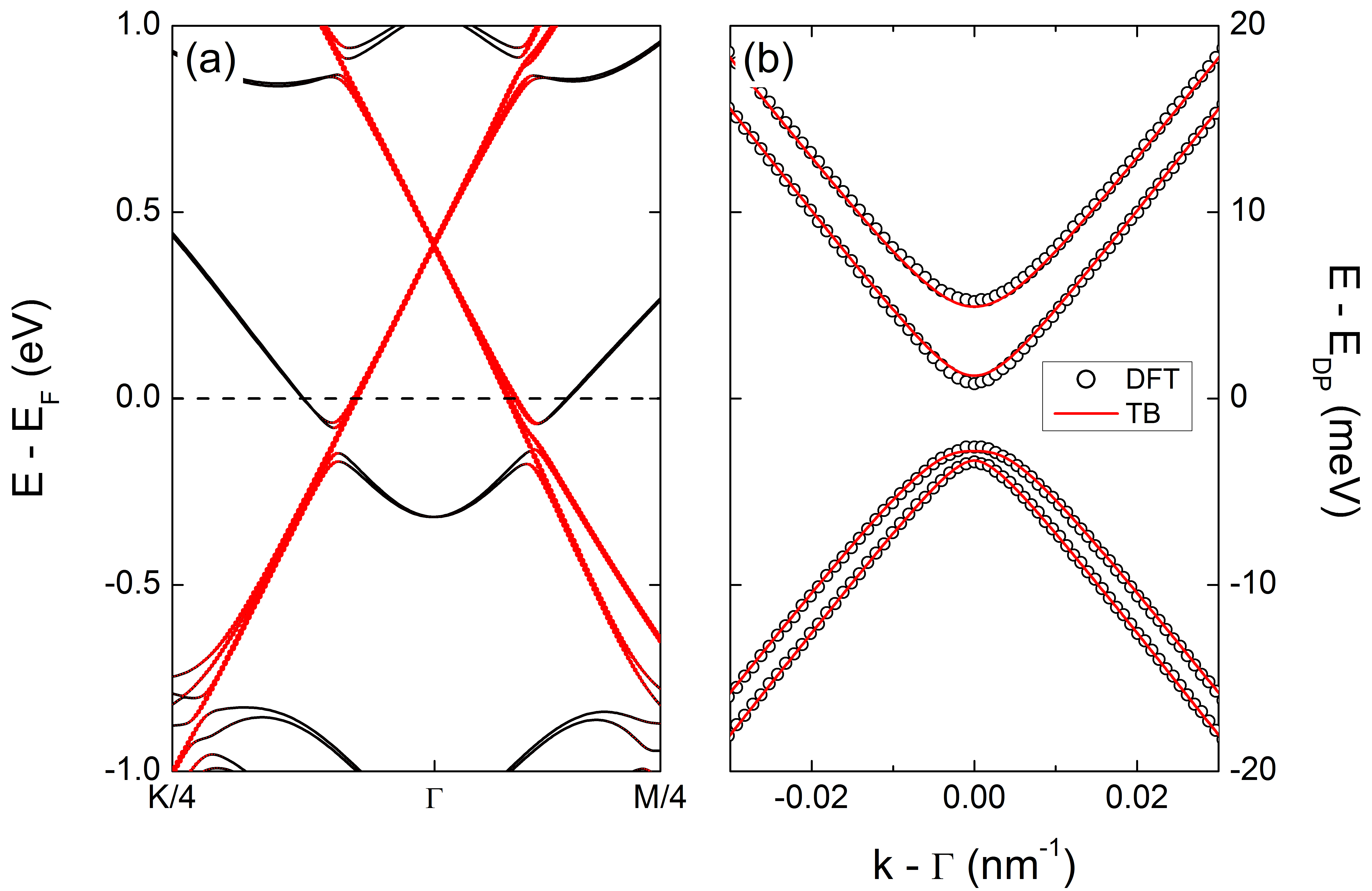}
\caption{Band structure of the Gr/Bi$_2$Se$_3$ heterostructure in the highly commensurate unit cell. (a) DFT band structure, where red symbols are the projection onto the carbon atoms. (b) Zoom of the graphene bands near the Dirac point, where symbols are the DFT results and lines are the fit using the TB model of Eq.\ (\ref{eq_tb}).}
\label{fig_bandstructure}
\end{figure}

To describe the electronic properties of graphene on a TI, we employ a TB Hamiltonian given by
\begin{align}
H &= \sum \limits_{\alpha=0,p} t_{\alpha} \sum \limits_{\langle ij \rangle,s}c_{is}^{\dagger}c_{js} \nonumber \\
&+ \frac{i}{3 \sqrt{3}} \sum \limits_{\langle\langle ij \rangle\rangle,ss'} c_{is}^{\dagger} c_{js'} (\lambda_I + \xi_i \lambda_{VZ}) [\nu_{ij} s_z]_{ss'} \nonumber \\
&+ \frac{2i}{3} \sum \limits_{\langle ij \rangle,ss'} c_{is}^{\dagger} c_{js'} [(\lambda_R^z \hat{z} + \lambda_R^\rho \hat{\rho}) \cdot (\mathbf{s} \times \mathbf{d}_{ij})]_{ss'} \nonumber \\
&+ \frac{2i}{3} \sum \limits_{\langle\langle ij \rangle\rangle,ss'} c_{is}^{\dagger} c_{js'} [(\xi_i \lambda_{PIA} \hat{z}) \cdot (\mathbf{s} \times \mathbf{D}_{ij})]_{ss'},
\label{eq_tb}
\end{align}
where $c_{is}^{\dagger}$ ($c_{is}$) is the creation (annihilation) operator of an electron at lattice site $i$ with spin $s$, $\mathbf{d}_{ij}$ ($\mathbf{D}_{ij}$) is the unit vector pointing from site $j$ to nearest (next-nearest) site $i$, $\mathbf{s}$ are the spin Pauli matrices, $\nu_{ij} = +1(-1)$ for a clockwise (counterclockwise) hopping path from site $j$ to $i$, $\xi_i = +1(-1)$ on sublattice A (B), and the single (double) brackets are sums over first (second) nearest neighbors. The first term in Eq.\ (\ref{eq_tb}) describes the hopping between nearest-neighbor carbon atoms. As depicted in Fig.\ \ref{fig_structure}(a), this has two different strengths: $t_0$ is the hopping within the carbon ring surrounding the top (green) Se atom, and $t_p$ is the hopping between carbon rings. This describes a Kekul\'e distortion of the graphene lattice in the hollow configuration, and opens a band gap of $2|t_0 - t_p|$ in the absence of SOC. In the large unit cell, Fig.\ \ref{fig_structure}(c), there is no Kekul\'e distortion and thus $t_0 = t_p$. The second term describes intrinsic SOC in the graphene lattice \cite{PhysRevLett.95.226801}, $\lambda_I$, and is assumed nonzero only for the carbon ring surrounding the top Se atom; this is highlighted by the solid triangle in Fig.\ \ref{fig_structure}(a). In the larger unit cell, this term is uniform. The third term is valley-Zeeman SOC, $\lambda_{VZ}$, which couples spin and valley and arises when sublattice symmetry is broken in the graphene layer \cite{Kochan2017}. For this reason, it is only present in the larger unit cell. The fourth term is a uniform Rashba SOC, $\lambda_R^z$, induced by an electric field perpendicular to the graphene plane. The fifth term is a second Rashba SOC, $\lambda_R^\rho$, arising from a radial in-plane electric field. We have found the best qualitative fit to DFT by choosing a nonuniform in-plane field, with nonzero values of $\lambda_R^\rho$ only along the green arrows in Fig.\ \ref{fig_structure}(a). In the larger unit cell this term does not exist, owing to the lack of radial symmetry. Finally, the last term is denoted PIA (pseudospin inversion asymmetry) SOC, $\lambda_{PIA}$, which is akin to a second-order Rashba SOC and leads to a $k$-linear spin splitting of the bands. This particular term only arises in the presence of sublattice symmetry breaking plus a perpendicular electric field \cite{Kochan2017}.

\section{Results}
We first focus on the highly commensurate structure, since this is the structure that has been exclusively studied in the literature up to now. Figure \ref{fig_bandstructure}(a) shows the DFT band structure of this system, where the red symbols denote the projection onto the carbon atoms. Since the Bi$_2$Se$_3$ layer is only 1QL thick, the surface states do not form, but they do appear in the 6QL structure (see Supplemental Fig.\ S1). In the Gr/Bi$_2$Se$_3$ heterostructure, there is significant charge transfer between the graphene and the TI, which induces a strong p-doping of the graphene and pushes its Dirac point into the TI conduction band. An equivalent amount of charge transfer is also seen in the bridge and top configurations of this unit cell. Because of the $\sqrt{3} \times \sqrt{3}$ supercell, the graphene Dirac cones are folded from the K and K$'$ points onto the $\Gamma$ point of the first Brillouin zone, resulting in eight nearly-degenerate bands. This degeneracy is broken by the Kekul\'e distortion and the SOC induced by the TI, resulting in a band gap opening and spin splitting of the conduction and valence bands. This can be seen in Fig.\ \ref{fig_bandstructure}(b), which shows a closeup of the graphene bands near the Dirac point. Black symbols are the DFT results and the red lines are the fits using the TB model of Eq.\ (\ref{eq_tb}). A band gap and spin splitting on the order of a few meV are observed. As mentioned above, each of these bands are nearly doubly degenerate due to the folding of K/K$'$ to $\Gamma$, but this degeneracy is broken by the Kekul\'e distortion and the SOC (see Supplemental Fig. S2).

\begin{figure}[t]
\includegraphics[width=\columnwidth]{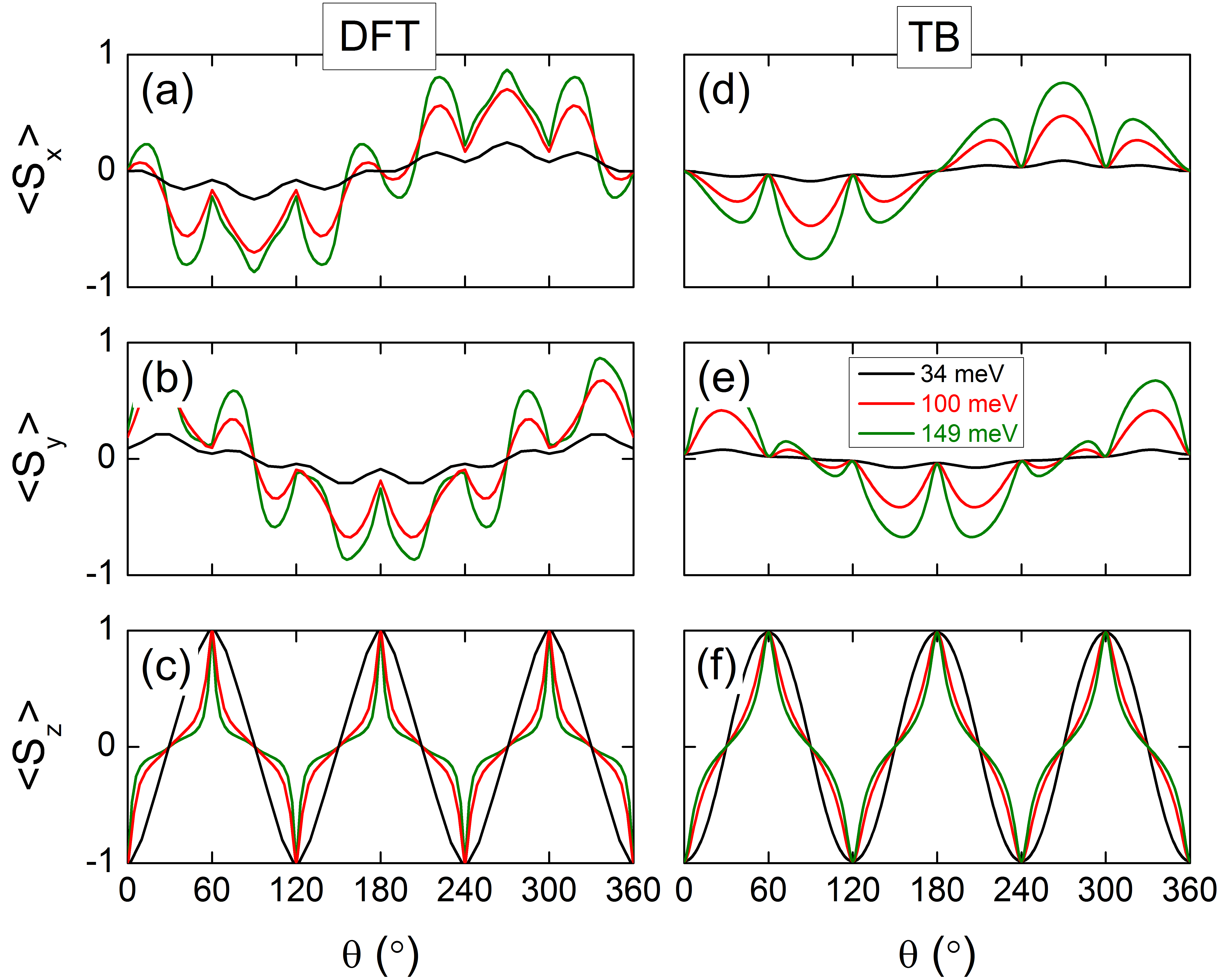}
\caption{Spin texture of Gr/Bi$_2$Se$_3$ in the highly commensurate structure. The left column (a)-(c) contains the DFT results and the right column (d)-(f) contains the TB results, showing each spin component as a function of angle $\theta$ around a constant energy contour. Black, red, and green curves are energy contours corresponding to 34, 100, and 149 meV above the graphene Dirac point.}
\label{fig_spintexture}
\end{figure}

Figure \ref{fig_spintexture} shows an overview of the spin texture of the graphene bands in the Gr/Bi$_2$Se$_3$ heterostructure. The top, middle, and bottom rows show the projections of the spin of the highest conduction band along the $x$, $y$, and $z$ axes respectively, plotted as a function of the angle $\theta$ around a constant energy contour. The first column contains the DFT results and the second column contains the TB fit. The black, red, and green curves indicate the energy dependence of the spin texture, at Fermi energies of 34, 100, and 149 meV above the graphene Dirac point. Several characteristic features of the spin texture can be seen. The first is that the $x$ and $y$ components exhibit an overall Rashba-like behavior, with $\langle S_x \rangle \sim -\sin \theta$ and $\langle S_y \rangle \sim \cos \theta$. However, this overall trend is punctuated by sharp minima every $60^{\circ}$. Second, the $z$ component of the spin is generally nonzero and shows maxima at these same points. These peaks correspond to points of anticrossing between the K and K$'$ bands that were folded to the $\Gamma$ point, which is enabled by the valley-mixing Kekul\'e distortion and the in-plane Rashba SOC (see Supplemental Fig.\ S2). Finally, the energy dependence shows that the in-plane components of the spin become weaker near the Dirac point, while the weight of the out-of-plane component increases.

Table \ref{table1} shows the TB parameters that best reproduce the DFT band structure and spin texture of the Gr/TI heterostructures, with the left column showing the case for the smaller unit cell. The orbital gap induced by the Kekul\'e distortion is $\sim$6 meV, and the SOC strengths are on the order of a few meV. A notable result is the relative magnitude of the in-plane and out-of-plane Rashba terms. Recent work has found good fits to the DFT band structure when assuming $\lambda_R^\rho \ll \lambda_R^z$ \cite{PhysRevB.87.075442}, but here to obtain the proper in-plane spin texture it is necessary to enforce $\lambda_R^\rho \gg \lambda_R^z$. An overview of the dependence of the in-plane spin texture on $\lambda_R^z$ and $\lambda_R^{\rho}$ can be found in Fig.\ S3 of the Supplemental Material. The intrinsic SOC $\lambda_I$ is necessary for a proper fit to the DFT band structure, but has no impact on the spin texture. Meanwhile, the out-of-plane spin component $\langle S_z \rangle$ depends crucially on the presence of the Kekul\'e distortion, which hybridizes the K and K$'$ bands; in its absence the magnitude of $\langle S_z \rangle$ drops by three orders of magnitude (see Supplemental Fig.\ S2). As stated above, the valley-Zeeman and PIA terms do not appear in this system because sublattice symmetry is not broken. Here we note that this spin texture is quite different from that in prior works, which predicted a purely Rashba-like behavior \cite{Beule2017, ZhangJunhua}. We attribute this difference to the choice of model used; the earlier works used a continuum model for the graphene and TI bands that does not account for trigonal warping, the Kekul\'e distortion, or the in-plane Rashba terms. We have found that both of these terms are crucial for properly capturing the DFT spin texture.

\begin{table}[t]
\begin{tabular}{c c c}
\toprule
Parameter & Small unit cell & Large unit cell \\
\midrule
$t_0$			& -2.6 eV	& -2.4 eV	\\
$t_0-t_p$			& -3		& 0		\\
$\lambda_I$		& -2.5	& 0		\\
$\lambda_{VZ}$		& 0		& -0.6	\\
$\lambda^z_R$		& 0.05	& 0.3		\\
$\lambda^{\rho}_R$	& -1.9	& 0		\\
$\lambda_{PIA}$	& 0		& -1.1	\\
\bottomrule
\end{tabular}
\caption{TB fits to the DFT band structure and spin texture of Gr/Bi$_2$Se$_3$ heterostructures. Unless otherwise specified, all quantities are in units of meV.}
\label{table1}
\end{table}

Figure \ref{fig_bigunitcell} shows the band structure and spin texture of the larger unit cell (see Fig.\ \ref{fig_structure}(c)), and the right column of Table \ref{table1} shows the TB fitting parameters. It is clear that, owing to the different interface symmetry of the larger unit cell, there are significant differences in the band structure, spin texture, and relevant SOC parameters compared to the smaller unit cell. In the band structure, the graphene Dirac cones remain separated at the K and K$'$ points of the Brillouin zone while the charge transfer between the graphene and the TI remains large (see Supplemental Fig. S6). Owing to this lack of band folding, hybridization between the valleys no longer occurs and the $60^\circ$ periodicity of the spin texture disappears. Instead, $\langle S_z \rangle$ remains independent of the momentum direction, and its sign is valley-dependent. This behavior is driven by the presence of valley-Zeeman SOC, $\lambda_{VZ}$, which is permitted by the sublattice symmetry breaking in the larger unit cell. The in-plane spin components follow the typical Rashba texture, and the PIA SOC determines their energy dependence. As mentioned above, the Kekul\'e distortion and in-plane Rashba SOC are not present in this system. Additionally, the intrinsic SOC is found to be vanishingly small. It it interesting to note that this TB model is identical to that used for Gr/TMDC systems, and the obtained fitting values are also quite similar \cite{Gmitra2016}.

\begin{figure}[t]
\includegraphics[width=\columnwidth]{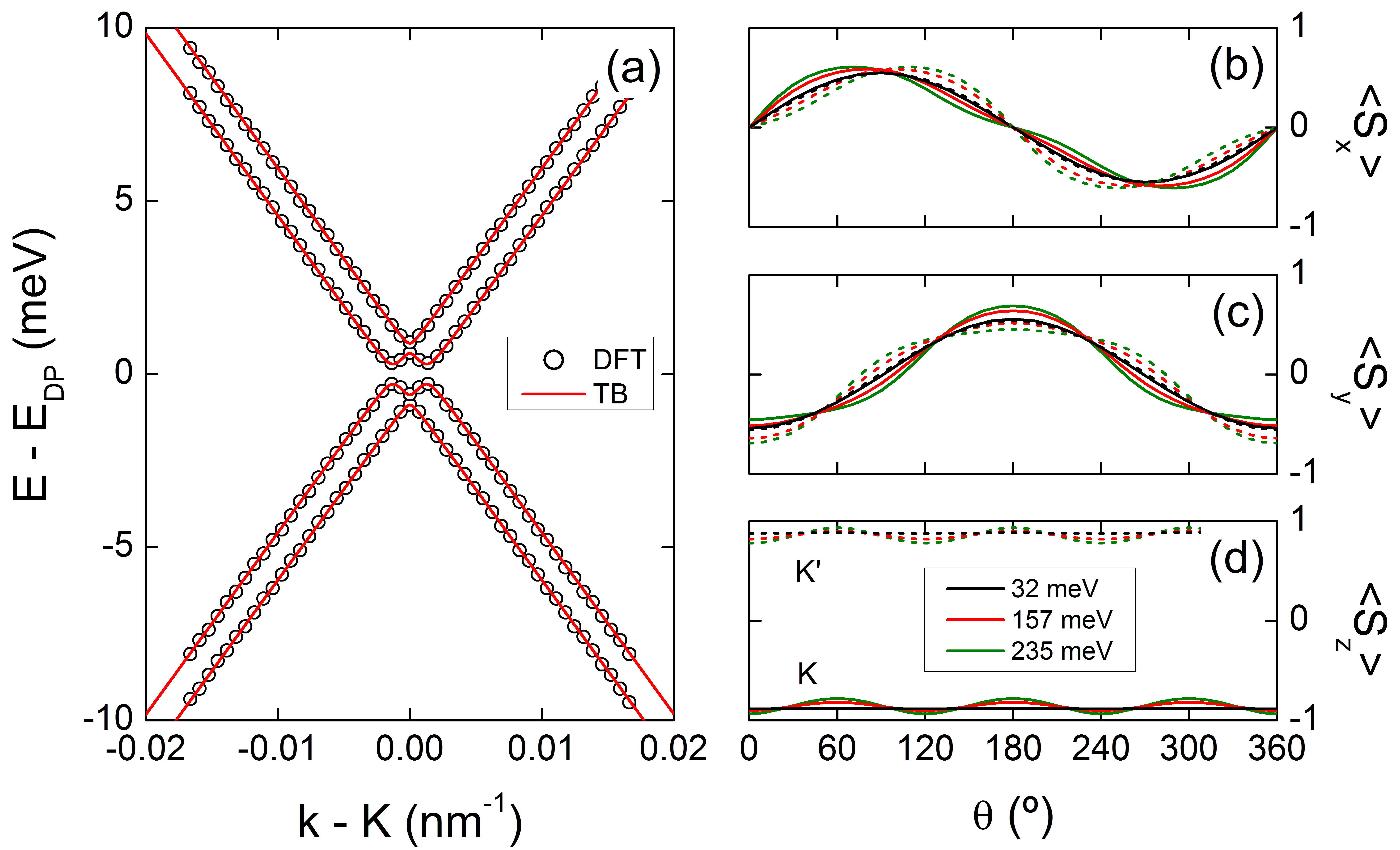}
\caption{Eletronic properties of Gr/Bi$_2$Se$_3$ in the large unit cell. (a) Band structure, where symbols are the DFT results and lines are the TB fit. (b)-(d) Spin texture from DFT, where black, red, and green curves are energy contours corresponding to 32, 157, and 235 meV above the graphene Dirac point. Solid (dashed) lines are for the K (K$'$) valley.}
\label{fig_bigunitcell}
\end{figure}

A comparison of Figs.\ \ref{fig_spintexture} and \ref{fig_bigunitcell} indicates that the symmetry of the system can have a significant impact on the spin texture induced in graphene by proximity to a TI substrate. This then raises the question: which of these scenarios is most likely to be encountered in an experimental setup? For experiments that interface graphene with TIs via stacking of exfoliated layers, the situation seen in the larger unit cell seems more likely, as the alignment and relative orientation of the graphene and TI lattices remains largely uncontrolled \cite{Herrero2015, Bian2016}. However, through careful processing and device fabrication, more precise control over the interface may be achieved \cite{Novoselov2016, Zhang2014}. From the perspective of spin transport, a measurement of the spin lifetime anisotropy can be invaluable in determining the dominant SOC terms and the nature of spin relaxation in these systems \cite{Raes2016, Raes2017}. Indeed, predictions of fast in-plane spin relaxation in Gr/TMDC heterostructures, driven by spin-valley locking and intervalley scattering \cite{Cummings2017}, have recently been confirmed by measurements of the spin lifetime anisotropy \cite{Ghiasi2017, Benitez2017}.

From our calculated spin textures we can predict the spin lifetime anisotropy of each system, defined as the ratio of out-of-plane to in-plane spin lifetime, $\zeta \equiv \tau_{s,z} / \tau_{s,x}$. Assuming the D'yakonov-Perel' regime of spin relaxation, the lifetime of spins polarized along $\alpha$ is given by $\tau_{s,\alpha}^{-1} = \tau_{\beta}^* (\overline{|\mathbf{\Omega}|^2} - \overline{\Omega^2_{\alpha}})$, where $\mathbf{\Omega}$ is the momentum-dependent effective magnetic field arising from SOC in units of spin precession frequency, $\tau_{\beta}^*$ is the time to randomize the $\beta$-component of $\mathbf{\Omega}$, with $\beta \perp \alpha$, and the overline represents an average over the Fermi surface at a particular Fermi energy \cite{Fabian2007}. For a given energy band, the effective magnetic field can be decomposed as $\mathbf{\Omega} = \omega \mathbf{S}$, where $\omega = \Delta E / \hbar$ is the spin precession frequency associated with the spin splitting $\Delta E$ of the band, and $\mathbf{S} = \langle \psi | \mathbf{s} | \psi \rangle$ is the spin polarization of the eigenstates $\psi$ associated with the band. The spin lifetime anisotropy arising from the spin-split band structure can then be written as
\begin{equation}
\zeta = \frac{\tau^*_z \sum \limits_{i=1}^{4}\left(\overline{|\mathbf{S}|^2} - \overline{S_x^2}\right)_{i}} {\tau^*_x \sum \limits_{i=1}^{4}\left(\overline{|\mathbf{S}|^2} - \overline{S_z^2}\right)_{i}},
\label{eq_anisotropy}
\end{equation}
where the sum over $i$ includes each of the four conduction or valence bands in the Fermi surface average. In the small unit cell, because both Dirac cones are folded to the $\Gamma$-point we have $\tau^*_x = \tau^*_z = \tau_p$, with $\tau_p$ the momentum relaxation time. In the large unit cell the graphene Dirac cones remain at K and K$'$, and owing to the presence of $\lambda_{VZ}$ we have $\tau^*_x = \tau_p$ and $\tau^*_z = \tau_{iv}$, where $\tau_{iv}$ is the intervalley scattering time \cite{Cummings2017}.

Figure \ref{fig_anisotropy} shows the spin lifetime anisotropy in the Gr/TI heterostructures, calculated from Eq.\ (\ref{eq_anisotropy}). Panels (a) and (b) are for the small and large unit cell respectively, the open circles show the DFT results, and the solid lines are from the TB fits. In the small unit cell, the anisotropy remains negligible away from the graphene Dirac point, on the order of $1/2$, and reaches values in the hundreds at the lowest energies. This trend also holds within the TI bandgap, where the graphene bands are completely in-plane and the anisotropy is 1/2. Such behavior results from the increase in weight of $\langle S_z \rangle$ near the Dirac point and the corresponding decrease of $\langle S_x \rangle$ and $\langle S_y \rangle$. The anisotropy obtained from the TB fit is both qualitatively and quantitatively similar to the DFT results. It should be noted, however, that the TB model does not fully account for all aspects of the behavior of $\langle S_z \rangle$. In particular, the DFT results show that the lowest pair of conduction and valence bands exhibit no out-of-plane spin texture, while the upper pair of conduction and valence bands show $\langle S_z \rangle$ similar to Fig.\ \ref{fig_spintexture}(c) (see Supplemental Fig.\ S4). Meanwhile, the TB results show identical magnitude of $\langle S_z \rangle$ for all bands; this leads to an overstimation of $\zeta$ by approximately a factor of two. 

Figure \ref{fig_anisotropy}(b) shows the anisotropy in the large unit cell, in units of $\tau_{iv} / \tau_p$. Both the TB fit and the analytical prediction derived for Gr/TMDC systems \cite{Cummings2017} show nice agreement with the DFT results. In this case, the anisotropy is characterized by a strong electron-hole asymmetry, which is driven by the relatively large value of $\lambda_{PIA}$; as shown in Ref.\ \citenum{Cummings2017}, the out-of-plane spin relaxation rate is proportional to $(a k_F \lambda_{PIA} \pm \lambda_R)^2$, where $a$ is the graphene lattice constant, $k_F$ is the Fermi wave number, and the +(-) is for the conduction (valence) band. At sufficiently negative energies, when $a k_F \lambda_{PIA} = \lambda_R$, this model predicts that the spin lifetime anisotropy will diverge.  In reality, when $\tau_s^\perp$ becomes sufficiently long another source of spin relaxation, such as contact dephasing or magnetic impurities, would take over, placing an upper bound on $\zeta$. In systems without a strong PIA SOC, the anisotropy would be independent of the Fermi energy. In the large unit cell, the anisotropy is driven by the SOC and the charge scattering through $\tau_{iv} / \tau_p$. In general, intervalley scattering is caused by structural defects such as dislocations, grain boundaries, vacancies, etc., as well as chemical adsorbates such as hydrogen, oxygen, or other hydrocarbons that could be deposited during device fabrication \cite{Roche2012}. Bi$_2$Se$_3$ is known to suffer from Se vacancies, which might also induce short-range Coulomb potentials and intervalley scattering in graphene \cite{Roche2012}. Measuring $\tau_p$ is straightforward, as it can be deduced from the mobility and charge density. For example, a typical carrier density of $2 \times 10^{12}$ cm$^{-2}$ coupled with a mobility of 6000 cm$^2$/V.s, as measured recently for a graphene/Bi$_2$Se$_3$ system \cite{ZhangACS2017}, yields $\tau_p\approx 100$ fs. Determining $\tau_{iv}$ requires a measurement of weak localization (WL), but in Gr/TI or Gr/TMDC systems the strong SOC leads to weak antilocalization (WAL), making it difficult to extract $\tau_{iv}$. So far, the best that has been done for a Gr/TMDC system is to measure WL in a region of the device that is not covered by the TMDC, and to assume that value as an upper bound of $\tau_{iv}$ in the Gr/TMDC region \cite{Yang2016}. We are not aware of any estimates of $\tau_{iv}$ in Gr/TI systems. Measurements of WL in graphene systems yield $\tau_{iv} / \tau_p$ in the range of 3-20 depending on the sample quality and Fermi energy \cite{Yang2016,Wu2007,Ki2008}. Very generally, $\tau_p$ is in the range of tens of fs and $\tau_{iv}$ is on the order of hundreds of fs to a few ps. Assuming $\tau_{iv} \approx 10\tau_p$ as a typical experimental situation, we would expect an anisotropy on the order of a few tens over the full range of gate voltage.

\begin{figure}[t]
\includegraphics[width=\columnwidth]{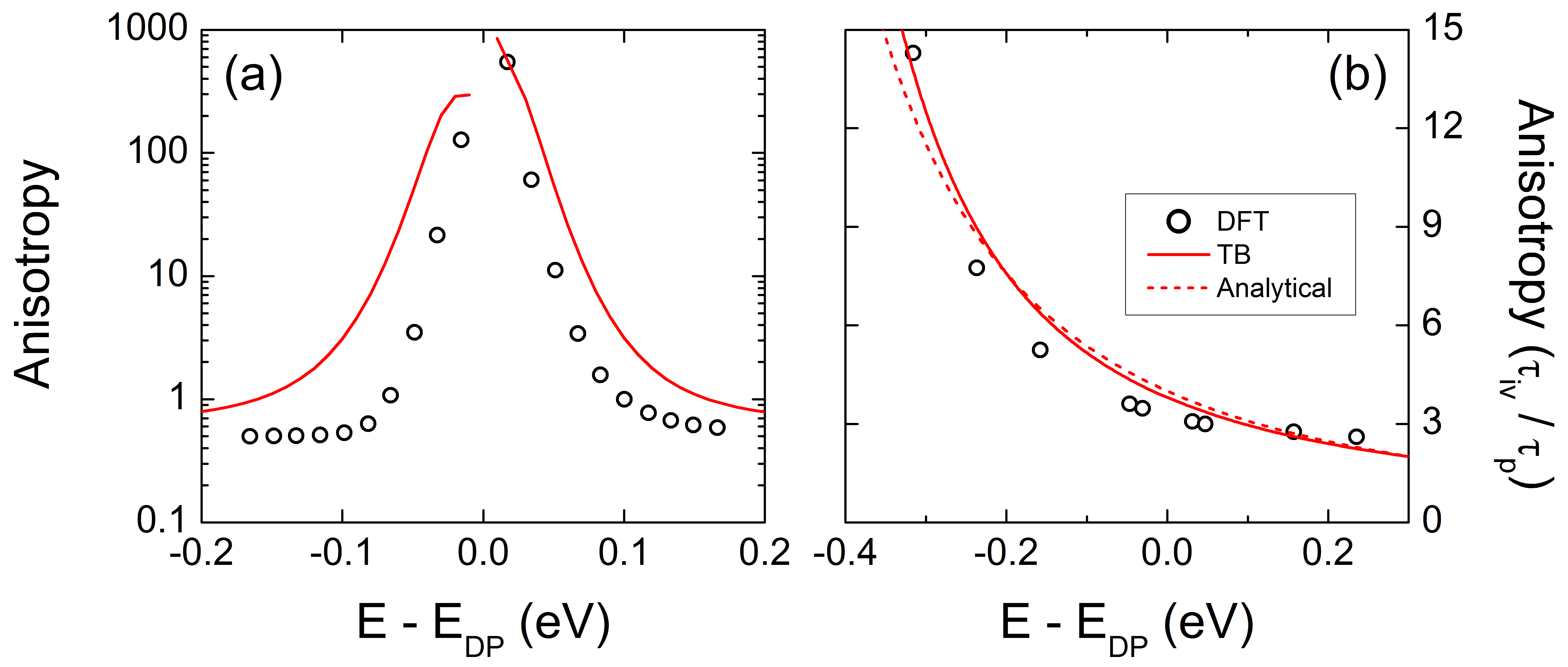}
\caption{Spin lifetime anisotropy of Gr/TI heterostructures as a function of energy relative to the graphene Dirac point. Panels (a) and (b) show the anisotropy in the small and large unit cell, respectively. Symbols are the DFT results, solid lines are the TB fits, and the dashed line is the analytical expression of Ref.\ \citenum{Cummings2017}.}
\label{fig_anisotropy}
\end{figure}

As mentioned in the Model and Methods section, we also used DFT to calculate the spin texture of the Gr/TI heterostructures in the top and bridge configurations, where the top Se atom sits below a carbon atom or a carbon-carbon bond, respectively. In contrast to the hollow configuration, these configurations show a small anisotropy, $\zeta \lesssim 1$ (see Supplemental Fig.\ S5). This arises from the lack of Kekul\'e distortion and/or in-plane Rashba SOC in these other lattice arrangements, which as discussed above, are both necessary to enhance the magnitude of $\langle S_z \rangle$ and thus the anisotropy.

\section{Conclusions} 
In summary, our study reveals the emergence of anisotropic spin transport in graphene in proximity with topological insulators, but the origin and energy dependence of this anisotropy vary significantly with the geometry of the interface. This arises from the very small lattice mismatch, which permits a highly commensurate unit cell at the appropriate twist angle. This is in contrast to the case of Gr/TMDC systems, which have a much larger lattice mismatch that precludes the formation of a small and highly commensurate unit cell \cite{Wang2015, Gmitra2016}. Similarly to the case of Gr/TMDC, Gr/TI displays an almost energy-independent anisotropy for zero twist angle between the graphene and TI lattices. However, in the highly commensurate unit cell (with a twist angle of $30^\circ$), the spin anisotropy is connected to both a Kekul\'e distortion and an in-plane Rashba SOC induced in the graphene by the TI. As a result, the spin lifetime becomes highly anisotropic near the graphene Dirac point while vanishing at higher energies, suggesting a much stronger variability via electrostatic gating in experiments. Such spin anisotropy could be playing a role in the debated experimental results reported to date in Gr/TI heterostructures \cite{VaklinovaNL2016,Zhang2016,ZhangACS2017}, while simultanously suggesting new device engineering such as gate-tunable linear spin polarizers, which remove the in-plane component of a spin-polarized current but leave the out-of-plane component intact.  

One useful observation is that, as shown in Fig.\ \ref{fig_bandstructure}, the Fermi level initially lies in the Bi$_2$Se$_3$ conduction band, which will generate parallel transport in the graphene and TI layers. However, given that the spin lifetime in the TI bulk should be exceptionally short (few femtoseconds) \cite{Zhang2013, Cummings2016}, any measured spin signal may still carry features of the spin transport in the graphene layer. To more optimally realize the conditions in which the TI surface states would play a role in the transport properties of Gr/TI heterostructures, ternary compounds, with the TI Fermi energy well within the TI bulk gap \cite{Xu2016}, would be even more desirable.

Finally, it would be interesting to include defects and disorder in the {\it ab initio} simulations and TB models, since this could locally alter the strength and nature of the SOC parameters. Such analysis, beyond the scope of the present work, could be also extended by developing a full Gr/TI tight-binding model, using for instance the Fu-Kane-Mele model \cite{Fu2007, Soriano2012}.

\begin{acknowledgement}
ICN2 is funded by the CERCA Programme / Generalitat de Catalunya, and is supported by the Severo Ochoa program from Spanish MINECO (Grant No.\ SEV-2013-0295). The authors acknowledge funding from the Spanish Ministry of Economy and Competitiveness and the European Regional Development Fund (Project No.\ FIS2015-67767-P MINECO/FEDER, FIS2015-64886-C5-3-P), the Secretar\'{i}a de Universidades e Investigaci\'{o}n del Departamento de Econom\'{i}a y Conocimiento de la Generalidad de Catalunya (2014 SGR 58, 2014 SGR 301), the European Union Seventh Framework Programme under grant agreement 696656 (Graphene Flagship), and the EU H2020-EINFRA-5-2015 MaX Center of Excellence (Grant 676598). DS thanks the Marie Curie NanoTRAINforGrowth Cofund program at INL.
\end{acknowledgement}

\providecommand{\latin}[1]{#1}
\makeatletter
\providecommand{\doi}
  {\begingroup\let\do\@makeother\dospecials
  \catcode`\{=1 \catcode`\}=2 \doi@aux}
\providecommand{\doi@aux}[1]{\endgroup\texttt{#1}}
\makeatother
\providecommand*\mcitethebibliography{\thebibliography}
\csname @ifundefined\endcsname{endmcitethebibliography}
  {\let\endmcitethebibliography\endthebibliography}{}

\end{document}